\documentclass[showpacs,preprintnumbers,amsmath,amssymb]{revtex4}

\usepackage{graphicx}
\usepackage{dcolumn}
\usepackage{bm}

\begin{document}

\preprint{NISHO-2006-1}

\title{Color Glass Condensates in dense quark matter \\and
quantum Hall states of gluons}

\author{Aiichi Iwazaki}
\affiliation{Department of Physics, Nishogakusha University, Ohi Kashiwa Chiba
  277-8585,\ Japan.}%
%



\date{April 30,2006}

\begin{abstract}
We apply the effective theory of color glass condensate
to the analysis of gluon states in dense quark matter,
in which the saturation region of gluons is also present.
We find that in the region  
two point function of gluons
shows algebraic long range order.
The order is completely the same as the one gluons
show in the dense quark matter, which form quantum Hall states.
The order leads to the vanishing of massless gluon
pole. We also find that the saturation region of gluons extends from small $x$ to
even large $x\lesssim 1$ in much dense quark matter.
We point out
a universality that the color glass condensate is
a property of hadrons at high energy and of quark matter at high baryon density.
\end{abstract}

\maketitle

\section{introduction}

It has recently been recognized both phenomenologically 
and theoretically\cite{rev} that color glass condensate (CGC)
is a realistic state of gluons in nucleons or nuclei.
The state is observed in deep inelastic scattering of nucleons
at high energy or high energy scattering between nuclei. Thus, it is considered to 
be associated with specific phenomena of QCD in such high energy scatterings.
But the idea is more general and applicable to
the other systems such as dense quark matter or quark gluon plasma. 
Here, we call the state of gluons whose number density is
saturated 
as the state of CGC,
although the use of the terminology might not be standard one\cite{rev}.   
The effective theory of CGC can clarify 
gluonic states, especially, their number density in both transverse $k_t$ 
and longitudinal $k^{+}$ momenta.
The theory is formulated by
using light cone formulation\cite{cone} in infinite momentum frame, $P^{+}\to \infty $.
( Usually, variable of $x\equiv k^{+}/P^{+}$ is used instead of $k^{+}$ 
where $P^{+}$ is the longitudinal momentum of the system itself under consideration.
Our notation is the standard one used in light cone formulation\cite{cone}. ) 
Then, only quantum effects of the leading order in $1/P^{+}$ are taken.
Thus, the CGC revealed by the theory is associated with the high energy scattering.
That is, gluons with much small $x$, which form the state of CGC,
can be seen only by deep inelastic scattering at high energy.
But, as we explain below, the region of CGC in dense quark matter 
extends from small $x\ll 1$ to even large $x\lesssim 1$. 
This implies that the CGC in the dense quark matter 
arises as a property of the matter 
in low energy physics just like
color superconducting phase of quark matter.

In this paper we apply the theory\cite{cgc} to dense quark matter, in which
the state of CGC is also present. The region of the CGC 
in standard $(k_t,1/x)$ plane extends 
from saturation momentum, $k_t=Q_s(x)$ to 
the vanishing momentum, $k_t=0$ 
because of quarks being deconfined in the dense quark matter. 
Since the saturation momentum, $Q_s^2(x)$, is proportional to 
the baryon number density of the quark matter, the CGC can be realized 
at not necessarily small $x$, e.g. $x=0.3$ for sufficiently dense quark matter. 
We discuss that in the region of 
the CGC, two point function of gluons shows an algebraic off-diagonal long range order
in the transverse directions with coordinate, $x_t$; 
$\langle tr(A(x_t)A(0))\rangle\to 1/(\alpha_s |x_t|^2)$ as $|x_t|\to \infty$.
It should be noted that the function is proportional to
$\alpha_s^{-1}$ in the CGC region; $\alpha_s\equiv g^2/4\pi$ stands for QCD fine structure constant. 
This behavior of the two point function in two dimensional transverse plane 
indicates the existence of a certain order in the CGC.
We show that exactly the same off-diagonal long range order of gluons as this one
is obtained in the dense quark matter where gluons
form a fractional quantum Hall state.
The quantum Hall state of the gluons have recently been discussed\cite{cf1,cf2,qhs} to be present
in dense quark matter where a color magnetic field is generated spontaneously:
under the color magnetic field gluons form quantum Hall states.
It is well known that a characteristic order is present in the states. 
Therefore, the CGC in the dense quark matter suggests the presence of the
quantum Hall state of gluons. On the other hand,
in the case of nuclei, as cyclotron radius of gluons is larger than the radius
of the nucleons, the quantum Hall states of gluons can not arise.
Thus, the CGC in the nuclei does not represent 
the quantum Hall states.
We have originally found the color ferromagnetic phase
in the laboratory frame of the quark matter, but, obviously
the quantum Hall states of gluons as well as
the existence of color magnetic field still remain 
in the infinite momentum frame; the direction of the magnetic field
is taken to point into $x_3=x^{+}-x^{-}$. 

In the next section \ref{2}, we evaluate the number density of gluons
by using the effective theory of the CGC. In the section \ref{3},
we show that the number density represents a $2$ point correlation
function of gluons, i.e. propagator, and that the correlation in the CGC region 
indicates the presence of an order in the dense quark matter.
In the section \ref{4}, we review quantum Hall states of gluons in color ferromagnetic
phase of the dense quark matter. In the section \ref{5},
we show that the same correlation as the 
one in the CGC is obtained in the quantum Hall states of gluons.
We discuss that the correlation in the CGC leads to the screenig
of color charge or the vanishing of gluon's massless pole.
This fact implies the presence of an order of the CGC.
In the section \ref{6}, we point out that the production process 
of the CGC is similar to that of gluon condensation 
in the dense quark matter, which form quatum Hall states.
We summarize our results in the section \ref{7}.

\section{\label{2}number density of gluons in quark matter}
\subsection{Effective theory of CGC}
In the present paper we consider $SU(N_c)$ gauge theory.
First, we define a number density of gluons in the dense quark matter using light cone formulation,
where in light cone gauge, $A^{+}=0$, 
the transverse gauge bosons, $A^{i}$($i=1.2$) is expanded in terms of
annihilation operators, 
$a^i(k^{+},k_t)$; $[a^i(p^{+},p_t),{a^j}^{\dagger}(k^{+},k_t)]=
\delta^{i,j}\delta(k^{+}-p^{+})\delta^{2}(k_t-p_t)$,

\begin{equation}
A^i(x^{-},x_t)=
\int_{k^{+}>0}\frac{dk^{+}d^2k_t}{\sqrt{(2\pi)^32k^{+}}}(a^i(k^{+},k_t)
\exp(ik^{+}x^{-}-ik_tx_t)+\mbox{h.c.})
\end{equation}
with transverse ( longitudinal ) momenta, $k_t$ ($k^{+}$) and coordinates, $x_t$ ( $x^{-}$ ).
We have not represented explicitly time coordinate, $x^{+}$ and color indices.
Then, the number density of gluons per unit transverse area is defined as

\begin{equation}
d^3N(x,k_t)/d^3k=\langle\mbox{matter}|
{a^{i}}^{\dagger}(k^{+},k_t)a^i(k^{+},k_t)|\mbox{matter}\rangle/(\pi R^2)
\end{equation}
where the summation over color indices is assumed implicitly.
$R$ denotes transverse radius of the matter which is supposed to be sufficiently large in order 
to have translational invariance in the transverse directions. We note that $R$ is independent of 
the baryon number density characterizing the dense quark matter, while it is proportional to $A^{1/3}$ where $A$ is the 
mass number of nuclei. The number density can be rewritten by using a field strength, 
$F^{i+}(x^{-},x_t)$,

\begin{eqnarray}
d^3N(\tau,k_t)/d\tau d^2k_t&=&k^{+}d^3N(x,k_t)/d^3k  \nonumber \\
&=&\int \frac{d^3x d^3y}{4\pi^4 R^2} \exp(ik(x-y))\langle
\mbox{matter}|F^{i+}(x)F^{i+}(y)|\mbox{matter}\rangle
\label{nd}
\end{eqnarray} 
with $\tau\equiv\log(1/x)$,
where $d^3x=dx^{-}d^2x_t$ and 
$k(x-y)\equiv k^{+}(x^{-}-y^{-})-k_t(x_t-y_t)$ and $F^{i+}(x)=-\partial^{+}A^{i}(x)$
in the light cone gauge. 

Now, we estimate the correlation function by using effective theory of CGC.
In order to do so,
we will explain briefly the theory.
First of all, it is assumed that there exist an effective theory including quantum effects of gluons and
quarks with longitudinal momentum from infinity to $\Lambda^{+}$
and with all transverse momentum. It is also assumed that
almost of all quantum effects of quarks are taken into account in the theory;
quantum effects of quarks with smaller momenta than $\Lambda^{+}$ make no
significant contributions. Then, the effective theory is described 
by standard Lagrangian of only gluons with momenta $k^{+}\leq\Lambda^{+}$
added with a color random source term,

\begin{equation}
\label{L}
L_{\Lambda}(A^{\mu},\rho)=-\int d^4x \frac{F^{\mu,\nu}F_{\mu,\nu}}{4}
+\frac{i}{gN_c}\int d^3x tr\Bigl(\rho(x^{-},x_t) 
P\exp(ig\int^{x^{+}}_{-\infty}dz^{+}A^{-}(z^{+},x^{-},x_t))\Bigl),
\end{equation}
where the color source, $\rho(x^{-},x_t)$, represents the quantum effects of quarks and gluons
with $k^{+}>\Lambda^{+}$,
which is supposed to have been integrated out in the original Lagrangian of QCD.
It is static ( independent of $x^{+}$ ) because it represents slowly moving components
of the quarks and gluons with momenta $k^{+}>\Lambda^{+}$.
It is only coupled with a gauge field, $A^{-}$ and it transforms as an ajoint representation
under gauge transformation. That is, the source describes modes propagating only in the positive $x^{+}$ direction
since we take infinite momentum frame.
We also note that the spatial extension in $x^{-}$
of the source is given by $(\Lambda^{+})^{-1}$
since it involves quantum effects with momentum, $k^{+}$, from infinity to $\Lambda^{+}$;
$\rho(x^{-},x_t)\neq 0$ only for $0<x^{-}<(\Lambda^{+})^{-1}$.
Using this Lagrangian, we find classical solutions, $A_{cl}(x,\rho)$ 
which depend on the source.
Then, the effective theory of CGC dictates 
that quantum expectation values are obtained
by averaging over the random source 
with a gauge invariant normalized distribution, $W_{\Lambda}(\rho)$; 
$\int D\rho W_{\Lambda}(\rho)=1$, 

\begin{equation}
\label{exp}
\langle\hat{O}(A(x,\rho))\rangle_{\Lambda}=\int D\rho \,\,O(A_{cl}(x,\rho))W_{\Lambda}(\rho)
\end{equation}
where $\hat{O}(A)$ represents an arbitrary operator of gauge fields and 
index $\Lambda $ indicates that the expectation value involves
only quantum effects with longitudinal momentum, $k^{+}$ such as $\infty>k^{+}>\Lambda^{+} $. 
$ D\rho$ is a gauge invariant measure.
Hence, the problem of obtaining the quantum averages is to find the distribution
function, $W_{\Lambda}(\rho)$ and to take the average with the use of the function.
All quantum averages over gluons and quarks are replaced by an average over the source 
with the use of $W_{\Lambda}(\rho)$. This simplification is the point of
the effective theory.

The classical static solution independent of $x^{+}$ is obtained easily in a 
transverse covariant gauge, $\partial_i A^{i}=0$ by putting $A^{-}=0$.
Then, the solutions are that $A^{\mu}(x^{-},x_t)=\delta^{\mu,+}\alpha(x^{-},x_t)$
where $-\partial^2_t \alpha(x^{-},x_t)=\rho_{\rm cv}(x^{-},x_t)$; 
$\partial^2_t\equiv\Sigma_{i=1}^2\partial^2_i$.
( The index ${\rm cv}$ of $\rho_{\rm cv}$ denotes the source in the covariant gauge. )
In order to obtain a solution in the light cone gauge, $A^{+}=0$,
we make a gauge transformation. Hence, it follows that 
in the gauge, $A^{+}=0$, $A^i(x^{-},x_t)=(i/g)V(x^{-},x_t)\partial^iV^{\dagger}(x^{-},x_t)$ 
with $V(x^{-},x_t)\equiv P\exp(ig\int^{x^{-}}_{-\infty}dz^{-}\alpha(z^{-},x_t))$
where $P$ denotes path ordering.  

We proceed to incorporate quantum effects of gluons with longitudinal momentum,
$b\Lambda^{+}<k^{+}\leq\Lambda^{+} $ with $0<b<1$. This is done by
integrating the gluon fields with the momenta. As a result we obtain 
a new effective theory with the same Lagrangian as the one in eq(\ref{L}) but 
with the distribution function, $W_{b\Lambda}(\rho)$, renormalized in which the
source recieves corrections and  
its spatial extension is modified such as
 $\rho(x^{-},x_t)\neq 0$ only for $0<x^{-}<1/(b\Lambda^{+})$.
The functional form of $W_{b\Lambda}(\rho)$ is also changed. 
Consequently, we obtain the distribution, $W_{y}(\rho)\equiv W_{b\Lambda}(\rho)$ 
at arbitrary longitudinal
momentum scale, $k^{+}=b\Lambda^{+}$; $y=\log(\Lambda^{+}/k^{+})=\log(1/b)$. 
Quantum expectation values at the scale of $b\Lambda^{+}$ can be 
obtained just as in eq(\ref{exp}) with the use of $W_{y}(\rho)$.

Instead of the distribution function, $W_y(\rho)$, of the source,
it is convenient to use the distribution function, 
$W_y(\rho_{\rm cv}=-\partial^2_t\alpha)$ of the gauge field, $\alpha$.
Hereafter we write it simply as $W_y(\alpha)$.
Jacobian associated with the change of variables, $\rho\to\alpha$,
is trivial. 

This $W_{y}(\alpha)$ obeys a renormalization group equation 
called as JIMWLK equation\cite{jim}, which is
a functional equation difficult to solve.
Instead of solving explicitly the equation,
we can derive a closed equation for an correlation function, 
$S_y(r_t)\equiv\frac{1}{N_c}\langle\mbox{matter}|tr(V^{\dagger}(r_t)V(0))|\mbox{matter}\rangle_y$
of a Wilson line $V(r_t)\equiv P\exp(ig\int^y_{-\infty} dz \alpha_z(x_t))$ 
where $\alpha_z(x_t)\equiv z\alpha(x^{-},x_t)$ with $z=\log(x^{-}\Lambda)$.
The equation is called as Balitsky-Kovchegov equation\cite{BK} which is obtained by
using JIMWLK equation along with taking both large $N_c$ limit and large 
baryon number density limit ( large mass number limit in the case of nuclei ),

\begin{equation}
\label{BK}
\partial_y S_y(r_t) =-\frac{N_c\alpha_s}{\pi}\int \frac{d^2z_t}{2\pi}\frac{r_t^2}{(r_t-z_t)^2z_t^2}
(S_y(r_t) -S_y(r_t-z_t)S_y(z_t)) \quad.
\end{equation}

With the use of $S_y(r_t)$, we can rewrite the number density of gluons,

\begin{equation}
\frac{d^3N(\tau,k_t)}{d\tau d^2k_t}=\frac{N_c}{2\pi^3g^2}\int d^2r_t e^{ik_tr_t}
\int_{-\infty}^{\tau} dy S_y(r_t)(-\partial_t^2)\partial_y\log(S_y(r_t)),
\end{equation}
where we have used an assumption that the correlation length in the longitudinal 
direction vanishes, i.e. $\langle\alpha_y(x_t)\alpha_z(u_t)\rangle\propto \delta(y-z)$.
This assumption is satisfied in the solutions\cite{W} of $W_y(\alpha)$ obtained in
a mean field approximation or a Gaussian approximation.  

\subsection{Saturation of gluon number density}
Consequently, the number density of gluons can be obtained by solving B-K equation (\ref{BK}). 
Our concern is the CGC region with
much large $\tau$ or much small $k_t$ and the asymptotic solution
relevant to the region has been found\cite{solution};
$S_y(r_t)\to \exp\Bigl(-c_0\bigl(\log(r_t^2Q_s^2(y))\bigl)^2\Bigl)$ 
for $|r_t|\gg (Q_s(y))^{-1} $ with a numerical constant, $c_0$.
Therefore, it follows that

\begin{equation}
\label{sat}
\frac{d^3N(\tau,k_t)}{d\tau d^2k_t} \simeq \frac{N_c^2-1}{16\pi^3g^2N_c}\log(Q_s^2(\tau)/k_t^2)
\quad \mbox{for} \quad k_t^2 \ll Q_s^2(\tau),
\end{equation}  
where $Q_s(\tau)$ is a saturation momentum.
On the other hand, it behaves roughly in the large $k_t$ such as
$\frac{d^3N(\tau,k_t)}{d\tau d^2k_t} \sim \frac{Q_s^2(\tau)}{k_t^2}
\quad \mbox{for} \quad k_t^2 \gg Q_s^2(\tau)$,
where an irrelevant constant is not explicitly written.
This equation implies that the number density grows very rapidly in 
the region of large $k_t$ with the decrease of the transverse momentum, $k_t$.
Similarly, it grows very rapidly with the decrease of longitudinal momentum, $x=\exp(-\tau)$
because $Q_s^2(\tau)\sim x^{-\eta}$
with $\eta\sim 0.3$\cite{solution}.

The behavior of the number density in small $x$ or small $k_t$ in eq(\ref{sat})
is quite different from the one in large $x$ or $k_t$.
The significant feature of
the formula in eq(\ref{sat}) is that the number density of gluons depends only on $k_t$ through the term, 
$\log(Q_s^2(\tau)/k_t^2)$ and depends on the gauge coupling constant, $g^2$ inversely.
That is, the number density of gluons saturates in both momenta
of $k_t$ and $x$; it grows only in logarithmic way because $Q_s^2(\tau)\sim x^{-\eta}$
with $\eta\sim 0.3$.
The saturated state of gluons is called as the state of CGC.
The feature is much intriguing and specific to QCD
or non Abelian gauge theories. It should be stressed
that the very feature gives rise to characteristic algebraic long range order of gluons
in two dimensional transverse plane discussed below.

\subsection{Saturation momentum in dense quark matter}
Here we would like to mention about the saturation momentum of the dense quark matter.
It may be defined in general such that $Q_s^2(y)=\alpha_sN_c\frac{1}{\pi R^2}\frac{dN}{dy}$.
We expect that gluon number density, $dN/dy$, is proportional to
the total baryon number of the system with the radius, $R$; $dN/dy\propto \rho_B R^3$
where $\rho_B$ is the normal nuclear density of nuclei or
the baryon number density of the quark matter. Thus,  
$Q_s^2(y)$ is proportional to $A^{1/3}$ in nuclei, while
it is proportional to the baryon number density in the quark matter.
We note that
the region of the CGC is given such that $x<x_c(k_t)$ for given $k_t$ 
where $x_c$ satisfies $k_t=Q_s(y_c(k_t))$ with $y_c(k_t)=\log(1/x_c(k_t))$.
Then, when we consider the quark matter with $\rho_B$ being sufficiently large,
the region of the CGC is achieved even 
with large $x\lesssim 1$.
This is because $Q_s^2(\tau)\sim \rho_B x^{-\eta}$ 
with $\eta\sim0.3$\cite{solution}.
For example, $Q_s(\rho_n,x=10^{-4})=Q_s(11 \rho_n,x=0.3)$ 
where $\rho_n\simeq 2.8\times 10^{14}\,\mbox{g/cm}^3$ stands for the
normal nuclear density. Namely, the saturation momentum of nuclei at small $x=10^{-4}$ 
is equal to the saturation momentum at $x=0.3$ of the quark matter 
with the baryon density, 
$11$ times larger than that of the nuclei. 
Therefore, the CGC of such dense quark matter is not neccessarily 
associated with high energy scattering. We can see the effects of the CGC
even in low energy scattering. 
Therefore, the CGC of hadrons is a feature of QCD observed at high energy limit,
while the CGC of quark matters is that of QCD observed at large baryon number density limit.

The result is naively suggested in the following
consideration. As has been recognized, the CGC can be analized perturbatively
because $Q_s(x)$ is sufficiently large at sufficiently small $x$ 
for $\alpha_s(Q_s)$ to be much less than $1$.
But, we need to take into account infinitely large number of gluons.
The analyis becomes non trivial. 
Similarly, the dense quark matter can be analized perturbatively 
because $\alpha_s(\mu_B)\ll 1$ with $\mu_B \gg \Lambda_{QCD}$ 
where $\mu_B$ is the chemical potential of the baryon number	;
$\rho_B\propto \mu_B^3$. As we explain below, we also need to
take into account infinitely large number of gluons for the analysis of dense quark matter. 
The analysis is non trivial and has been done in the laboratory frame
of the quark matter. In this sence, the CGC in the dense quark matter 
may arise even at low energy scattering.

\section{\label{3}long range order of gluons in cgc}
Now, we rewrite the number density in eq(\ref{nd}) 
in a different way. 
Since $\langle\mbox{matter}|F^{i+}(x)F^{i+}(y)|\mbox{matter}\rangle=0$ for $x^{-}\neq y^{-}$
in the effective theory as we have stated,
we can integrate over $x^{-}$ and $y^{-}$ by putting $\exp(ik^{+}(x^{-}-y^{-}))=1$.
Then, it follows that using the translational invariance in the transverse direction, 

\begin{equation}
d^3N(\tau,k_t)/d\tau d^2k_t=\frac{1}{4\pi^2}\int d^2r_t 
\exp(-ik_tx_t)\langle\mbox{matter}|A^i_{\tau}(x_t)A^i_{\tau}(0)|\mbox{matter}\rangle
\end{equation}
where we have used the fact that $A^i(x^{-},x_t)=0$ for $x^{-}\leq 0$ and have denoted 
$A^i_{\tau}(x_t)\equiv A^i(x^{-},x_t)$ for $x^{-}>1/k^{+}$; 
$\tau=\log(1/x)=\log(\Lambda^{+}/k^{+})+\log(P^{+}/\Lambda^{+})$.
Note that since $\alpha(x^{-},x_t)=0$ for $x^{-}>1/k^{+}$,
$A^i_{\tau}(x_t)$ is independent of $x^{-}$ when $x^{-}>1/k^{+}$.

Therefore, we find that the number density of gluons is just the propagator of
the gluons in two dimensional transverse space. In other words, the number density
describes the two point correlation function of gluons. Consequently, we can find
an intriguing property that the correlation function shows 
an algebraic off-diagonal long range order in 
the region of CGC,

\begin{equation}
\label{co}
\langle\mbox{matter}|A^i_{\tau}(x_t)A^i_{\tau}(0)|\mbox{matter}\rangle= c/(g^2 x_t^2) \quad \mbox{for}
\quad |x_t| > Q_s^{-1}(\tau),
\end{equation}
where $c$ is a numerical constant independent of quark number density, or $Q_s(\tau)$.
It should be noted that we can take $|x_t|$ larger than 
the confinement scale $\Lambda_{QCD}^{-1}$
in the dense quark matter because quarks are deconfined.

Although the behavior of the correlation at large distance is not
usual one showing a long range order 
such as $\langle \phi(x_t)\phi(0) \rangle\to \langle\phi(r_t)\rangle\langle\phi(0)\rangle\neq 0$
as $|x_t|\to\infty$, the algebraic decay of the correlation function in two dimensional spaces 
implies the existence of an order in the system.
Actually,
the algebraic decay in two dimensional spaces 
usually arises in the correlations of phases such as
for example, $\langle\exp(-i\theta(x_t))\exp(i\theta(0))\rangle\propto 1/|x_t|^{\beta_e}$ 
in the low temperature phase of XY model; the variable of $\exp(i\theta(x_t))$ represents 
the direction of spin.
When we write the spin variable 
such as $S(x_t)=n(x_t)\exp(i\theta)$, then the spin shows the algebraic long range order,
$\langle S(x_t)S(0)\rangle\propto 1/|x_t|^{\beta_e}$, 
but $\langle n(x_t)n(0)\rangle\neq 0$ as $|x_t|\to\infty$
because of the magnitude, $n$, of the spin being fixed.
Another example is correlations of composite bosons, $\phi$, 
in fractional quantum Hall states\cite{qhs,nboson}
of two dimensional electrons
with filling factor, $\nu=2\pi n_e/eB$ ;
$\langle\phi(x_t)\phi(0)\rangle\propto 1/|x_t|^{1/2\nu}$ as $|x_t|\to \infty$ where
$n_e$ is the number density of electrons.
In this case, the algebraic decay arises from the phase of the field, $\phi(x_t)$.
It is well known that this algebraic decay of the correlation
implies the existence of an order in the quantum Hall states of electrons.
This order characterizes the quantum Hall states\cite{qhs,iwa}.

\section{\label{4}quantum hall states of gluons}
We have recently shown that the dense quark matter possesses a color ferromagnetic phase\cite{cf1,cf2}
in which gluons form a quantum Hall state 
under a color magnetic field generated spontaneously.
The phase is realized in the region of baryon number density
being sufficiently large for the perturbation theory or loop expansions to be valid.
But it arises
in the lower baryon density
than the baryon density necessary for the realization of color superconductivity\cite{color}.
Thus, the phase is more important phenomenologically, e.g. 
for neutron star physics\cite{magnetar}.
We will show below that this quantum Hall state of gluons with $\nu=1/4$ also 
exhibits the same off-diagonal long range order as the one in eq(\ref{co}) 
as well as $1/g^2$ dependence. ( In the present section, we do 
use the standard formulation of gauge theories instead of the light cone formulation. 
Our main concern is quantum Hall states of gluons
realized in two dimensional plane, which we identify as the transverse plane
in the above discussion. Correlations of gluons discussed below hold
even in the light cone formulation. )

The presence of color magnetic field makes lower one loop effective potential\cite{savidy}
of gluons. This is perturbatively reliable result in much dense quark matter,
since the gauge coupling constant is sufficiently small in the matter.
( Even if we take into account the free energy of quarks,
the minimum of the effective potential is determined mainly by gluons loops.
Thus, the field strength determined in this way has only the slight dependence of
quark chemical potential. )  
Although the presence of the field minimizes the one loop potential, 
there still exist more stable states, which are beyond the one loop approximation. 
Those are obtained by taking account of the repulsive $4$ point interaction of gluons;
the one loop approximation does not address 
with the $4$ point interaction.
That is, they are 
the quantum Hall states of
gluons coupled with the field.
Actually, it is easy to see in the SU(2) gauge theory 
that under the color magnetic field, ${\cal B}\propto \sigma_3$,  
gluons with color $\sigma_{1,2}$ have such energy spectra as 
$E_{\pm}^2(n,k_3)=2g{\cal B}(n+1/2)+k_3^2\pm 2g{\cal B}$
with integer $n\geq 0$ specifying Landau levels; $k_3$ is a momentum
parallel to the field, ${\cal B}$. Here, we take spatial ( color ) direction of ${\cal B}$ be pointed  
into $x_3$ ( $\sigma_3$ ) direction. 
Thus, the gluons with $E_{-}^2(n=0,k_3)=k_3^2-g{\cal B}<0$ for $k_3<\sqrt{g{\cal B}}$ are 
unstable\cite{unstable} and their amplitudes grow unlimitedly. In other wards
they are produced unlimitedly
if there is not the $4$ point interaction.
Especially, most unstable gluons ( they are modes with $k_3=0$ ) are two dimensional ones 
since there is no dependence of the coordinate,$x_3$. 
Their amplitudes grow unlimitedly, but eventually they are saturated to
form the quantum Hall state owing to the repulsive 4 point interaction of gluons.
This situation is very similar to the production process of gluons with small $x$ 
in CGC. That is, the gluons with smaller $x$, are produced unlimitedly by the gluons 
with larger $x$ but they are saturated due to the nonlinear
interaction of gluons.
Anyway, in the dense quark matter a diagonal color magnetic field ( ${\cal B}\propto \sigma_3$ )
is generated
spontaneously and off-diagonal gluons $\propto \sigma_{1,2}$ are produced
to form quantum Hall states. 
Explicitly, the gluons forming the quantum Hall state are given by 
$\phi_{\rm u}=(\Phi_1-i\Phi_2)\sqrt{\ell/2}$ where
$\Phi_{i}=(A_{i}^1+iA_{i}^2)/\sqrt{2}$ with spatial indices, $i$. 
A constant, $\ell$, appearing for the normalization of the field, $\phi_{\rm u}$,
denotes the spatial extension
in $x_3$ direction of a region ( quantum well ) in which
quantum Hall states of the gluons are fabricated;
the scale is determined by the other unstable modes with the momentum, $k_3$ such as
$\sqrt{g{\cal B}}>k_3>0$.
Thus, it is order of the magnetic length, $1/\sqrt{g{\cal B}}$. 

In order to describe the quantum Hall states of the gluons, $\phi_{\rm u}$,
it is convenient to use composite
gluons, $\phi_a$ instead of using $\phi_{\rm u}$; 
gluons attached with fictitious Chern-Simons flux, $a_{\mu}$\cite{cf1,cf2,sem,iwa}.
Using the gluon composite field, $\phi_a$, 
Lagrangian in three space-time dimensions for describing the quantum Hall state
is given by 

\begin{equation}
\label{la}
L_a=|(i\partial_{\nu}-gA_{\nu}+a_{\nu})\phi_a|^2+2g{\cal B}|\phi_a|^2-\frac{\lambda}{2}|\phi_a|^4+
\frac{\epsilon^{\mu\nu\lambda}}{4\alpha}a_{\mu}\partial_{\nu}a_{\lambda},
\end{equation}
with $\lambda=g^2/\ell$ and $A^{\nu}=(0,-{\cal B}x_2/2,{\cal B}x_1/2)$ 
representing color magnetic field ${\cal B}$,
where $\nu$ runs from $0$ to $2$ and the statistical factor $\alpha$
should be taken as $\alpha=2\pi\times $integer for the field $\phi_a$ to 
represent the composite boson. Antisymmetric tensor is such that $\epsilon_{\mu\nu\lambda}=1$ for
$(\mu,\nu,\lambda)=(0,1,2)$, otherwise $\epsilon_{\mu\nu\lambda}=0$. 
Quantization of the system leads to the following Hamiltonian,

\begin{equation}
\label{H}	
H(\phi_a,\vec{a})=P_a^{\dagger}P_a+|(i\vec{\partial}+\vec{a}-g\vec{A})\phi_a|^2
-2g{\cal B}|\phi_a|^2+\frac{\lambda}{2}|\phi_a|^4,
\end{equation} 
with commutation relations, 
$[\phi_a(x_t),P_a^{\dagger}(y_t)]=[\phi_a^{\dagger}(x_t),P_a(y_t)]=\delta^2(x_t-y_t)$,
where we have used a gauge condition, $a_0=0$.
Additionally, we have a constraint such that $\epsilon^{ij}\partial_ia_j(x)=2\alpha j(x)$
where $j(x)=i(\phi_a^{\dagger}P_a-P_a^{\dagger}\phi_a)$ is color charge density.
Canonical momentum of $\phi_a$ is given by $P_a=(\partial_0-igA_0+ia_0)\phi_a^{\dagger}$.
Quantum Hall states are obtained by finding groundstate solutions of the Hamiltonian.
Especially, in the mean field approximation we can easily obtain the spatially uniform
classical solution minimizing
$H$, which represents quantum Hall states. That is, by putting $g\vec{A}=\vec{a}$
and by introducing Lagrange multiplier, $b_0$ for the constraint, 
the classical solution is given\cite{cf2} by solving the following equations,

\begin{equation}
2b_0|\phi_a|^2=\frac{g{\cal B}}{2\alpha} \quad \mbox{and}\quad
b_0^2+2g{\cal B}=\lambda |\phi_a|^2.
\end{equation}  
The solution is that 
$|\phi_a^{cl}|\simeq \sqrt{2g{\cal B}/\lambda}=\sqrt{2{\cal B}/g\ell}$ 
for small gauge coupling $g^2\ll 1$. Quantum Hall states are 
characterized as
the condensed states of the boson, $\langle\mbox{qhs}|\phi_a|\mbox{qhs}\rangle=\phi_a^{cl}$
in the composite boson formulation.

Here we point out that 
the relation between the original gluon variable, $\phi_{\rm u}$ and 
the composite one, $\phi_a$ is given by\cite{sem}

\begin{equation}
\phi_{\rm u}(x)=\exp(i\int d^2y\theta(x-y)j(y))\phi_a(x),
\end{equation}
where $\theta(x)$ denotes azimuthal angle of $x$.
With the use of the relation, we obtain Hamiltonian of the original gluon field, $\phi_{\rm u}$
with the replacement such as $H(\phi_a\to\phi_{\rm u},a_{\mu}=0)$ in eq(\ref{H}).
Hence, Hamiltonian of the original gluon, $\phi_{\rm u}$, is given by

\begin{equation}
\label{Hu}
H=P_{\rm u}^{\dagger}P_{\rm u}+|(i\vec{\partial}-g\vec{A})\phi_{\rm u}|^2-2g{\cal B}|\phi_{\rm u}|^2+
\frac{\lambda}{2}|\phi_{\rm u}|^4,
\end{equation}
which can be obtained by extracting only the mode, $\phi_{\rm u}$, from the original Hamiltonian of gluons. 
We note that the Hamiltonian looks like the one of a Higgs model.
This implies that the groundstate of the Hamiltonian is formed by
a condensation of the gluons, $\phi_{\rm u}$. Such a condensation is explicitly realized
in the composite boson formulation.

\section{\label{5}long range order of gluons in dense quark matter}
Instead of $\phi_a$ we may introduce a different form\cite{nboson} of a composite boson, 
$\phi_b\equiv \exp(-i\int^t dt'a_0(t',x_t))\phi_a$ in the Lagrangian, then, it follows that

\begin{equation}
L_a=|i\partial_0\phi_b|^2-|(i\partial_i-gA_i+a_i-P_i)\phi_b|^2
+2g{\cal B}|\phi_b|^2-\frac{\lambda}{2}|\phi_b|^4+
\frac{\epsilon^{\mu\nu\lambda}}{4\alpha}a_{\mu}\partial_{\nu}a_{\lambda},
\end{equation}
with $P_i\equiv \partial_i\int^t dt'a_0(t',x_t)$.
Using this Lagrangian, we find that longitudinal $P_i$ 
and transverse $a_i$ 
are canonical conjugate with each other because of the last term 
$\int d^3x \frac{1}{2\alpha}\epsilon_{ij}\partial_t P_i a_j$ in $\int d^3xL_a$.
Actually, we can rewrite the last term such that 
$\int d^3x \frac{\epsilon^{\mu\nu\lambda}}{4\alpha}a_{\mu}\partial_{\nu}a_{\lambda}=
\int d^3x\frac{1}{2\alpha}(\partial_0a-\partial_0P)\vec{\partial}^2b$ where we put $P_i=\partial_iP$
and $a_i=\partial_ia+\epsilon^{ij}\partial_jb$. Thus, changing the variable, $P\to P+a$,
we find that only transverse component, $b$, of $a_i$ remains in the Lagrangian.
Consequently, we may regard that $a_i$ is transverse. 
Then, quantizing the canonical conjugate variables, 
$\delta a_i=a_i+gA_i$ and $P_i$ ( more precisely, $P$ and $-\vec{\partial}^2b/2\alpha$
are canonical conjugate ),
the term of Hamiltonian, 
$|(i\partial_i-gA_i+a_i-P_i)\phi_b|^2\sim |\langle\mbox{qhs}|\phi_b|\mbox{qhs}\rangle|^2(\delta a_i-P_i)^2$,
can be diagonalized where $|\langle \mbox{qhs}|\phi_b|\mbox{qhs}\rangle|^2=|\phi_a^{cl}|^2$.
Then, it is straightforward to show in a similar way to the one in ref.\cite{nboson} that

\begin{equation}
\langle\mbox{qhs}|\exp(-i\int^tdt'a_0(t',x_t))\exp(i\int^t dt''a_0(t'',0))|\mbox{qhs}\rangle
\to (\frac{1}{|x_t|\Lambda})^{\alpha/2\pi}
\quad \mbox{as}\quad |x_t|\to \infty,
\end{equation}
where we have noted that $P=\int^t dt'a_0(t',0)$ and that 
the quantum Hall state is the groundstate of the diagonalized Hamiltonian.
$\Lambda$ denotes a typical scale in QCD,
not necessarily equal to $\Lambda_{QCD}$ 
although it is proportional to $\Lambda_{QCD}$.
Therefore, we find that

\begin{equation}
\langle\mbox{qhs}|\phi_b^{\dagger}(x_t)\phi_b(0)|\mbox{qhs}\rangle\to 
|\phi_a^{cl}|^2(\frac{1}{|x_t|\Lambda})^{\alpha/2\pi}
=(2{\cal B}/g\ell)(\frac{1}{|x_t|\Lambda})^{\alpha/2\pi}
\quad \mbox{as} \quad |x_t|\to \infty,
\end{equation}
or,
\begin{equation}
\label{cor}
\langle\mbox{qhs}|\phi^{\dagger}_{\rm u}(x_t)\phi_{\rm u}(0)|\mbox{qhs}\rangle\to 
(2{\cal B}/g\ell)(\frac{1}{|x_t|\Lambda})^{\alpha/2\pi}
\quad \mbox{as} \quad  |x_t| \to \infty,
\end{equation}
where we have used the fact that $\int\ d^2x \,\theta(x)\langle j\rangle=0$,  
since the color charge distribution, $\langle j\rangle\equiv \langle \mbox{qhs}|j(x)|\mbox{qhs}\rangle$, in
the quantum Hall state is
uniform and spherical symmetric.
Hence, it turns out that the correlation function in quantum Hall state
of gluons shows the same algebraic
off-diagonal long range order as the one in eq(\ref{co}) of the color glass condensate
when its filling factor, 
$2\pi\langle \mbox{qhs}|j(x)|\mbox{qhs}\rangle/g{\cal B}=\pi/\alpha$, is equal to $1/4$ ( $\alpha=4\pi$ ).
We should note that since $g{\cal B}$ is given by $\Lambda_{QCD}^2$
in the one loop approximation\cite{savidy,unstable}
and $\Lambda \propto \Lambda_{QCD}$, the correlation function eq(\ref{cor})
behaves such that 
$\langle \phi_{\rm u}^{\dagger}(x_t)\phi_{\rm u}(0)\rangle\sim 1/(g^2|x_t|^2)$ as $|x_t|^2\to \infty $. 
The dependence of $g^2$ is the specific feature in eq(\ref{co})
characterizing the color glass condensate.

Therefore, we find that the off-diagonal long range order, eq(\ref{co}) of the gluons in CGC
is completely the same as the one in eq(\ref{cor}) of the gluons 
forming the fractional quantum Hall state
in dense quark matter. This similarity is stressed by rewriting eq(\ref{co})
with the use of $\phi_{\rm u}$ in the case of SU(2) gauge theory,

\begin{equation}
\langle\mbox{matter}|A^i_{\tau}(r_t)A^i_{\tau}(0)|\mbox{matter}\rangle  
=\langle\mbox{matter}|(\phi^{\dagger}_{\rm u}(r_t)\phi_{\rm u}(0))/l|\mbox{matter}\rangle+
\langle\mbox{matter}|A^i_{\tau,3}(r_t)A^i_{\tau,3}(0)|\mbox{matter}\rangle, 
\end{equation}
where $A^i_{\tau,3}$ is a gauge field in $\sigma_3$ direction of color space.
Furthermore, we can argue that if the algebraic decay of the correlation,
$\langle\mbox{matter}|A^i_{\tau}(r_t)A^i_{\tau}(0)|\mbox{matter}\rangle$,
comes from the gluon condensation, $\langle\phi_{\rm u}\rangle\neq 0$, 
of quantum Hall states as discussed above,
the color magnetic field, ${\cal B}=F^{1,2}_3$ should be present.

\begin{equation}
\langle F^{1,2}_3\rangle \sim\partial^1\langle A^2_3\rangle-\partial^2\langle A^1_3\rangle
+g(\langle A^1_1\rangle \langle A^2_2\rangle -\langle A^1_2\rangle \langle A^2_1\rangle)
\sim \Lambda^2/g \simeq {\cal B}
\end{equation}
since $g{\cal B}=\Lambda^2_{QCD}$ and $\Lambda\propto\Lambda_{QCD}$.

On the other hand, the color magnetic field, $F^{1,2}$, vanishes exactly in the CGC because of
the absence of a color source current pointing to the transverse direction 
as a leading term
in $1/P^{+}$. Thus, we may speculate based on the above discussion that
non-leading terms in $1/P^{+}$ would generate the magnetic field, $F^{1,2}$.

As has been pointed out\cite{screening,rev},
the color source, $\rho$ is screened completely in the region of the CGC;
$\int d^2x_td^2y_t\langle\rho( x_t )\rho ( y_t )\rangle=0$ and $\langle\int d^2x_t\rho\rangle=0$.
This originates from the long range behavior of the two point function in eq(\ref{co}) 
such as $1/|x_t|^2$. ( Strictly speaking, the property of the screening holds only in
the dense quark matter because we need to take the limit of $|x_t|$ infinity. 
In the case of hadrons, $|x_t|$ must be less than the confinement scale, $\Lambda^{-1}_{QCD}$.
On the other hand, the screening becomes complete only in the limit of $|x_t|$ infinity
because of the absence of explicit screening length. ) 
The fact of the screening can be stated in different way. Namely, the long range
behavior leads to the disappearance 
of the massless pole in the gluon propergater; 
$k_t^2\,(d^3N(\tau,k_t)/d\tau d^2k_t)\to 0$ as $k_t^2\to 0$.
The screening or the disappearance of the massless pole indicates the presence of an order in the system.
It apparently seems to be a Debye screening in color charged gluon gas.
But it is not correct. Because the effective theory of the CGC dictates 
that $F^{-i}=0$, that is,
color electric field, $\vec{{\cal E}}$
is identical to color magnetic field, $\vec{{\cal B}}$ ( ${\cal E}^1={\cal B}^2$ 
and ${\cal E}^2=-{\cal B}^1$ ),
the electric screening implies simultaneously the magnetic screening.  
So the screening is not simply a Debye screening.
The magnetic screening usually arises in superconducting states
where Meissner mass is generated. On the other hand, 
there is no such mass generation in the CGC because the correlation
function does not decay exponentially. Thus, the CGC in dense quark matter does not
implies color superconductivity\cite{color}.
Although there is not an explicit mass generation,
the screening indicates the existence of an order in this system. 
We speculate that the order is the one in quantum Hall states of gluons.

\section{\label{6}quantum hall state and cgc}
Finally, we point out another similarity between the CGC and quantum Hall state of gluons.
Namely, the formation process of the CGC is very similar to 
that of gluon condensation in quantum Hall state, which is described by a Higgs model\cite{cf2}. 
When the population of gluons with large $x$ is small, $N\ll 1$,
the population grows very rapidly ( BKFL region\cite{BKFL} ) due to the production
of gluons with smaller $x$. But when the population 
becomes large enough, $N\sim 1$, to interact with each other as $x\to 0$, the nonlinear interaction
becomes effective to make the population be saturated ( CGC region ).
Such a situation may be described\cite{population} by an equation of population dynamics,
$\partial_tN=N-N^2$ with an initial condition, $N(t=0)\ll 1$. Here, "time", $t$, 
is qual to the rapidity, $y=\log(1/x)$.
We note that 
BK equation in eq(\ref{BK}) governing the gluon density has a similar structure to this one.

A similar equation can be derived in a Higgs model just like
the Hamiltonian in eq(\ref{Hu}) of $\phi_{\rm u}$ 
describing quantum Hall states of unstable gluons. 
Although the relevant Hamiltonian in eq(\ref{Hu}) is not a typical Higgs model which 
we use below, the essence in the evolution of the gluon field can be simulated in the model.
That is,
from the equation of motion for the spatially uniform Higgs field,
$\partial_t^2\phi=\mu^2\phi-\frac{\lambda}{2}|\phi|^2\phi$,
we can derive an equation for the square of amplitude, $N_h\equiv|\phi|^2$, of the Higgs field,

\begin{equation}
\partial_tN_h=2\mu N_h\sqrt{1-\frac{\lambda N_h}{4\mu^2}}\simeq 2\mu N_h-\frac{\lambda N_h^2}{4\mu} 
\quad \mbox{when}\quad \lambda \ll 1,
\end{equation}
with the use of an condition, $\partial_tN_h=0$ at $N_h=0$.
We have assumed the phase of Higgs field being static, for simplicity.
Then, we find that when the amplitude of Higgs field is small ( $N_h\ll \mu^2/\lambda$ ), 
the state is unstable
and its amplitude grows exponentially in time. But when the amplitude becomes sufficiently large
( $N_h\sim \mu^2/\lambda$ ), the field saturates and stops to grow 
due to the $4$ point self interaction, $\lambda |\phi|^4$.
This similarity suggests that the formation of the CGC in nuclei or quark matter
may be understood in terms of a Higgs model in the above sense.

\section{\label{7}conclution}
We conclude that the long range behavior of the two point function of gluons 
in the CGC of dense quark matter 
is the same as that of gluons forming quantum Hall states.
The behavior leads to the screening of the color source or the disappearance of 
the massless pole of gluons. Hence, it
indicates the presence of an order in the state of the CGC
of the dense quark matter, just like the order in the quantum Hall states. 
Furthermore, the formation process of the CGC has a similarity 
to that of the quantum Hall states of gluons.
We may speculate that the CGC in dense quark matter might be just a condensate of gluons 
forming the quantum Hall states.
We have also shown that the saturation region of the CGC extends 
from small $x\ll 1$ to large $x\lesssim 1$, e.g. $x=0.3$.
The fact implies a universality of the CGC which is realized in 
both hadrons at high energy scattering and
quark matters at high baryon density. 
The universality may be applicable for quark gluon plasma
at high temperature.

\vspace*{2em}
We would like to express thanks
to Prof. O. Morimatsu, Dr. T. Nishikawa
and Dr. M. Ohtani for useful discussion.
We also express thanks to Dr. K. Itakura for useful comments.
This work was supported by Grants-in-Aid of the Japanese Ministry
of Education, Science, Sports, Culture and Technology (No. 13135218).


\newpage
\begin{thebibliography}{99}
\bibitem{rev}E. Iancu, A. Leonidov and L. McLerran, hep-ph/0202270.\\
E. Iancu and R. Venugopalan, hep-ph/0303204.\\
K. Itakura, hep-ph/0511031.
\bibitem{cone}for a review, see T. Heintl, hep-th/0008096.
\bibitem{cgc}L.D. McLerran and R. Venugopalan, Phys. Rev. D49 (1994) 2233;
Phys. Rev. D49 (1994) 3352; Phys. Rev. D50 (1994) 2225.
\bibitem{cf1}A. Iwazaki and O. Morimatsu, Phys. Lett. B 571 (2003) 61; 
A. Iwazaki, O. Morimatsu, T. Nishikawa and M. Ohtani,
Phys. Lett. B 579 (2004) 347.
\bibitem{cf2}A. Iwazaki, O. Morimatsu, T. Nishikawa and M. Ohtani,
Phys. Rev. D 71 (2005) 034014.

\bibitem{qhs}{\it The Quantum Hall Effect, 2nd Ed.}, edited by R.E. Prange
  and S.M. Girvan ( Springer-Verlag, New York, 1990 ).\\
S. Das Sarma and A. Pinczuk (Eds.),
Perspectives in Quantum Hall Effects, (Wiley, New York, 1997).
\bibitem{jim}J. Jalilian-Marian, A. Kovner, A. Leonidov and H. Weigert, Phys. Rev. D59 (1999) 014014. \\
E. Iancu and A. Leonidov and L.D. MacLerran, Nucl. Phys. A692 (2001) 583.
\bibitem{BK}I. Balitsky, Nucl. Phys. B463(1996) 99.\\
Y. Kovchegov, Phys. Rev. D60 (1999) 034008.
\bibitem{W}E. Iancu, K. Itakura and L. McLerran, Nucl. Phys. A724 (2003) 181.
\bibitem{solution}D. Triantafyllopoulos, Nucl. Phys. B648 (2003) 293.\\
E. Iancu, K. Itakura and L. McLerran, Nucl. Phys. A708 (2002) 327.
\bibitem{nboson}R. Shankar and G. Murthy, Phys. Rev. Lett. 79 (1997) 4437. 
\bibitem{color}K. Rajagopal and F. Wilczek, hep-ph/0011333.
\bibitem{magnetar}A. Iwazaki, Phys. Rev. D72 (2005) 114003.
\bibitem{savidy}G.K. Savvidy, Phys. Lett. B 71 (1977) 133.\\
H. Pagels, Lecture at Coral Gables, Florida, 1978.
\bibitem{unstable}N.K. Nielsen and P. Olesen, Nucl. Phys. B 144 (1978) 376;
Phys. Lett. B 79 (1978) 304.
\bibitem{sem}G.W Semenoff, Phys. Rev. Lett. 61 (1988) 517.
\bibitem{iwa}Z.F. Ezawa, M. Hotta and A. Iwazaki, 
Phys. Rev. B46, 7765 (1992);
Z.F. Ezawa and A. Iwazaki, J. Phys. Soc. Jpn. 61, 4133 (1990).
\bibitem{screening}A.H. Mueller, Nucl. Phys. B643 (2002) 501.
\bibitem{BKFL}E.A. Kuraev, L.N. Lipatov and V.S. Fadin, Sov. J. Phys. JETP 45 (1977) 199.\\
I.I. Balitsky and L.N. Lipatov, Sov. J. Nucl. Phys. 28 (1978) 822.
\bibitem{population}K. Itakura, hep-ph/0410336, proceedings for ICHEP2004.
\end{thebibliography}
\end{document}